\documentstyle[aps]{revtex}

\def\eeq{\end{equation}}
\def\beq{\begin{equation}}
\def\bea{\begin{eqnarray}}
\def\eea{\end{eqnarray}}
\begin{document}

\title{Aging in coherent noise models and natural time}
\author{Ugur Tirnakli\thanks{%
e-mail: tirnakli@sci.ege.edu.tr}}
\address{Department of Physics, Faculty of Science, 
Ege University, 35100 Izmir, Turkey}
\author{Sumiyoshi Abe\thanks{suabe@sf6.so-net.ne.jp}}
\address{Institute of Physics, University of Tsukuba, 
Ibaraki 305-8571, Japan}

\maketitle

\begin{abstract}
Event correlation between aftershocks in the coherent noise 
model is studied by making use of natural time, which has recently been 
introduced in complex time-series analysis. It is found that the aging 
phenomenon and the associated scaling property discovered in the observed 
seismic data are well reproduced by the model. It is also found that 
the scaling function is given by the $q$-exponential function appearing 
in nonextensive statistical mechanics, showing power-law decay of 
event correlation in natural time.

\noindent
{\it PACS Number(s): 89.75.Da, 05.65.+b, 05.90.+m, 91.30.-f}
\end{abstract}


\vspace{1.5cm}

In recent years, there has been an increasing interest in extended 
dynamical systems exhibiting avalanches of activity, whose size 
distribution is scale-free. Examples of such systems are 
earthquakes \cite{earth}, rice piles \cite{sand}, extinction in 
biology \cite{ext}, evolving complex networks \cite{nets}, and so on. 
Up to now, there is no unique and unified theory for such systems, 
but one of the candidates may be the notion of self-organize 
criticality (SOC) \cite{soc}. A key feature common in SOC models is that 
the whole system is under the influence of a small driving force that acts 
locally. These systems evolve towards a critical stationary state having 
no characteristic spatio-temporal scales without fine-tuning parameters. 
Two celebrated examples of such extended driven dynamical systems are 
the Olami-Feder-Christensen model of earthquakes \cite{ofc} and the 
Bak-Sneppen model for biological evolution \cite{bs}.

On the other hand, there exists another kind of simple and robust 
mechanism producing scale-free behavior in the absence of criticality. 
An important example is the coherent noise model \cite{newman1}, which we 
shall study here. The coherent noise model has been introduced to describe 
large-scale events in evolution. It is based on the notion of external 
stress coherently imposed on all agents of the system under consideration. 
Since this model does not contain any direct interaction among agents, 
it does not exhibit dynamical criticality. Nevertheless, it yields a 
power-law distribution of event (i.e., avalanche) size $s$, which is defined 
by the number of agents that change their states at each time step. 
The model allows existence of aftershocks. This is a direct consequence of 
the fact that, in the coherent noise model, the probability of a large 
event to occur is increased immediately after a previous large event.

The coherent noise model is defined in a rather simple manner. 
Consider a system, which consists of $N$ agents. Each agent $i$ has a 
threshold $x_i$ against external stress $\eta$. The threshold levels are 
chosen at random due to some probability distribution $p_{thresh}(x)$. 
The external stress is also chosen randomly due to another distribution 
$p_{stress}(\eta)$. An agent becomes eliminated if it is subjected to the 
stress $\eta$ exceeding the threshold for the agent. In practice, 
dynamics of the model can be summarized as the following three steps: 
(i) at each time step, a random stress $\eta$ is generated from 
$p_{stress}(\eta)$ and all the agents with $x_i\le \eta$ become eliminated 
and are replaced by new agents with new thresholds drawn from 
$p_{thresh}(x)$, 
(ii) a small fraction $f$ of the $N$ agents should be chosen at random and 
given new thresholds, and then (iii) go back to (i) for the next time 
step. Here, (ii) corresponds to the probability for the $f$ fraction of 
the whole agents of undergoing spontaneous transition. This is necessary 
for preventing the model from grinding to a halt 
(see e.g. Ref. \cite{wilkeD} for details).

The coherent noise model has been applied to the problems of 
earthquakes \cite{wilkeD,newman2}. There, the Gutenberg-Richter law for 
the relation between frequency of the events and the values of the 
magnitude \cite{GRlaw} and the Omori law for temporal decay pattern of 
aftershocks \cite{omori} are found to be realized. In particular, 
the Omori law, which is of relevance to our subsequent discussion, 
states that the rate of the number of aftershocks, $\rho$, after a 
mainshock at $t=0$ obeys

\beq
\rho \sim t^{-\tau}\;\; .
\eeq

\noindent
Empirically, the exponent $\tau$ takes a value between $0.6$ and $1.5$. 
This describes the slow power-law decay, and each of the relevant parts of 
the seismic time series is nonstationary. Such a time interval is 
referred to as the Omori regime.

Quite recently, the physical properties of correlation in the 
seismic time series have been studied in \cite{abe} based on analysis 
of the observed seismic data. It was shown that the aging phenomenon 
occurs inside the Omori regime but it disappears outside. The definite 
scaling property has also been identified for the event correlation 
function. These results suggest that nonstationarity of the Omori 
regime may have something to do with glassy dynamics.

A point of crucial importance regarding the aging phenomenon of 
earthquake aftershocks is that, unlike ordinary discussions of the 
aging phenomenon, the two-point correlation function is defined in the 
domain of "natural time" \cite{time1}, not conventional time. 
The concept of natural time is a kind of an {\it internal clock} counting 
the discrete event number. It has successfully been applied to revealing 
physical essence of complex time series such as seismic electric signals, 
ionic current fluctuations in membrane channels and so on \cite{time1}. 
The fact that natural time is more fundamental than conventional continuous 
time is still empirical. In this respect, one may recall that the concept 
of continuum is recognized through continuous physical processes, 
whereas in the case of earthquakes one is concerned with a series 
of discrete events. However, clearly, more investigations are needed 
for deeper understanding of natural time.

Here, we discuss the physical properties of the Omori regime in 
the coherent noise model. We shall see that aging and scaling 
discovered in the observed seismic data are well reproduced by the 
model. The correlation function is found to have the form of the 
$q$-exponential function (see below), showing slow power-law decay. 
These results have also striking similarities with those recently 
obtained for the nonextensive Hamiltonian system of an infinite-range 
coupled rotors in the course of nonequilibrium relaxation 
dynamics \cite{tamarit} in conventional time, although the systems 
as well as the chosen random variables are completely different from 
each other. This, in turn, indicates that the aging/scaling phenomena 
may be quite universal in extended dynamical systems out of equilibria, 
no matter if critical or not.

We have carried out the numerical study of the aging phenomenon 
in the coherent noise model with the exponential distribution for 
the external stress

\beq
p_{stress}(\eta) = a^{-1} \exp\left(-\frac{\eta}{a}\right) \;\;\;\; 
(a>0)\;\; ,
\eeq

\noindent
and the uniform distribution $p_{thresh}(x)$ $(0\le x \le 1)$ for the 
threshold level. The results obtained are in order.

First of all, we present in Fig.~1 a typical subinterval of 
the obtained time series of activity, where aftershocks are clearly 
identified. In Fig.~2, the rate of the probability of finding aftershocks 
larger than $s_1$ following the initial large event at $t=0$ is plotted 
with respect to the elapsed time $t$. The straight line represents the 
Omori law, which allows us to identify the Omori regime. We have 
ascertained that this result is insensitive to the threshold value $s_1$.

To investigate the property of event correlation, following the 
idea proposed in \cite{abe}, we have employed as the basic random 
variable the time of the $n$th aftershock with an arbitrary avalanche 
size, $t_n$, where $n$ is aforementioned natural time in the setting 
of our problem. The two-point correlation function is given by

\beq
C\left(n+n_w,n_w\right) = \frac{\left<t_{n+n_w} t_{n_w}\right> - 
\left<t_{n+n_w}\right> \left<t_{n_w}\right>}
{\left(\sigma_{n+n_w}^2 \sigma_{n_w}^2\right)^{1/2}} \;\; ,
\eeq

\noindent
where the average is understood as the ensemble average taken in 
association with a number of numerical runs, and the variances in the 
denominator are given by, 
$\sigma_m^2 = \left<t_m^2\right>-\left<t_m\right>^2$. 
It is noticed that this averaging procedure is different from that 
in \cite{abe}, in which the time average is used. 
Since the Omori regime is nonstationary, the correlation function 
depends not only on $n$ but also on $n_w$. In Fig.~3, we present the plots 
of $C\left(n+n_w,n_w\right)$ for several values of 
"natural waiting time" $n_w$. There, the clear aging phenomenon can be 
appreciated. Furthermore, as shown in Fig.~4, collapse of these curves 
can nicely be realized, following the scaling relation

\beq
C\left(n+n_w,n_w\right) = \tilde{C}\left(\frac{n}{n_w^{\alpha}}\right)\;\; ,
\eeq

\noindent 
where $\tilde{C}$ is a scaling function and $\alpha$ is numerically 
$\alpha\simeq 1.05$. This form of scaling is remarkably similar to the 
one presented in \cite{abe}, where the observed seismic data are employed.

To find the form of the scaling function, we have examined the 
semi-$q$-log plot of the curve in Fig.~4, where the $q$-logarithmic 
function is defined by

\beq
\ln_q(x) = \frac{x^{1-q}-1}{1-q} \;\;\;\;\; (x>0) \;\; ,
\eeq

\noindent 
which is the inverse function of the $q$-exponential function

\beq
e_q(x) = \left[1+(1-q)x\right]_+^{1/(1-q)}
\eeq	

\noindent
with the notation $[z]_+ = \max\{0,z\}$. (In the limit $q\rightarrow 1$, 
$\ln_q(x)$ and $e_q(x)$ converge to the ordinary logarithmic and exponential 
functions, respectively.) This pair of functions is known to play a 
central role in nonextensive statistical mechanics \cite{tsa1,tsa2}. 
The straight line of the correlation function in the semi-$q$-log plot 
shown in Fig.~5 indicates that the scaling function is given by the 
$q$-exponential. Therefore, event correlation decays slowly, following a 
power law. This reminds us of the recent discussion \cite{tamarit} about 
the nonextensive Hamiltonian system in the course of nonequilibrium 
relaxation, though the time variable employed there is the conventional one.

In conclusion, we have studied the physical properties of event 
correlation in the Omori regime of aftershocks in the coherent 
noise model. We have found that aging and scaling in natural time 
discovered in the observed seismic data can be reproduced remarkably 
well by this model. We have also found that the scaling function is 
given by the $q$-exponential function, and thus event correlation decays 
according to a power law.

\section*{Acknowledgments}
U.T. is supported by the Turkish Academy of Sciences, 
in the framework of the Young Scientist Award Program 
(UT/TUBA-GEBIP/2001-2-20). 
S. A. thanks Dr. H. Tanaka for discussions about natural time. 
He also thanks the financial support of the Grant-in-Aid for 
Scientific Research of Japan Society for the Promotion of Science. 

\vspace{1.5cm}


{\bf Figure Captions}

\vspace{1cm}

Figure~1: A section of the time series of activity in the 
dimensionless units. $5000$ time steps are shown after the largest 
shock out of the time series of $30000000$ events. 
The aftershocks following the mainshock at $t=0$ are 
clearly recognized.\\

Figure~2: A histogram of the time distribution (i.e., the rate) 
of $90000$ aftershocks larger than $s_1=10$ following the mainshock at 
$t=0$. The strength of the external stress in eq.(2) is $a=0.001$, and 
the fraction $f=5\times 10^{-6}$ of $N=200000$ agents is chosen at random. 
It obeys a power law in eq.(1) with $\tau\simeq 1$. 
All quantities are dimensionless.\\

Figure~3: Dependence of the event correlation function 
$C\left(n+n_w,n_w\right)$ of the aftershocks larger than $s_1=1$ on 
natural time. The strength of the external stress in eq.(2) is $a=0.2$ 
and the fraction $f=0.01$ of $N=10000$ agents is chosen at random. 
The ensemble average over $120000$ realizations is performed. 
The values of natural waiting time are 
$n_w=250$, $500$, $1000$, $2000$ and $5000$ from bottom to top. 
All quantities are dimensionless.\\

Figure~4: Data collapse for the correlation function 
$C\left(n+n_w,n_w\right)$ shown in Fig.~3.
The gray solid line corresponds to $e_q\left(-0.7 n/n_w^{1.05}\right)$ 
with $q\simeq 2.98$.\\

Figure~5: The semi-$q$-log plot of the collapsed curve with 
$n_w=1000$, $2000$ and $5000$ in Fig.~4. 
The straight line shows that the scaling function is of the 
$q$-exponential form with $q\simeq2.98$.

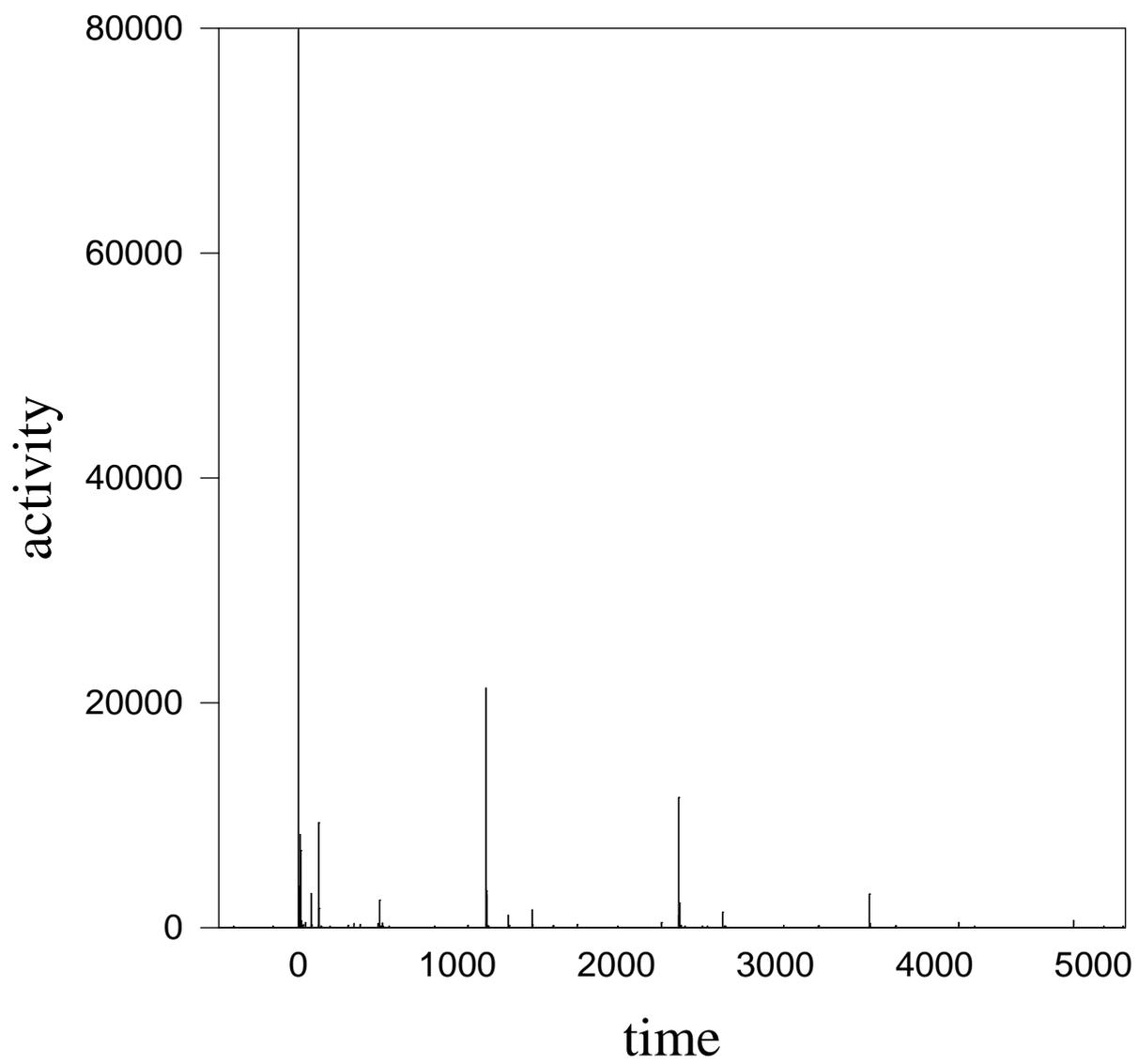
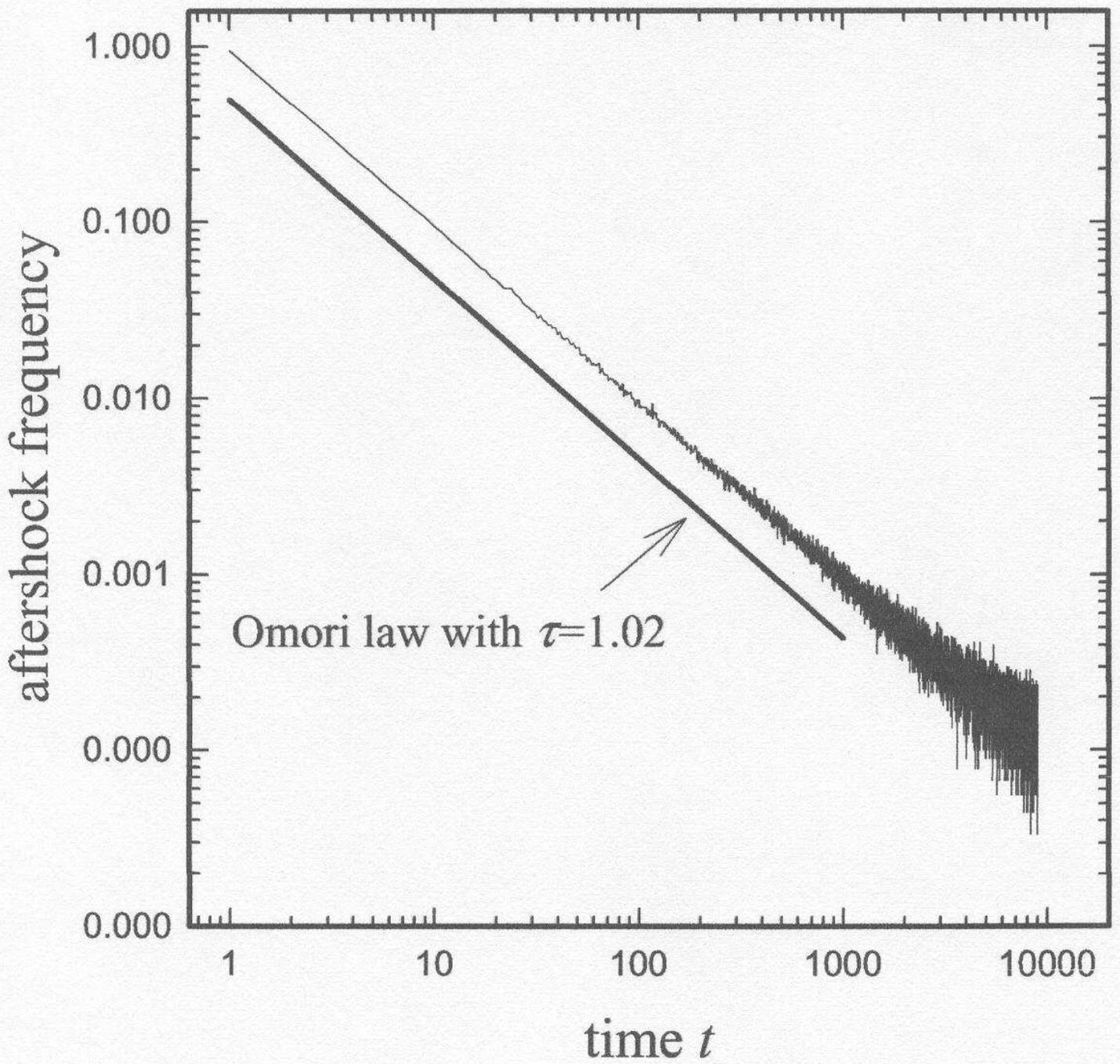
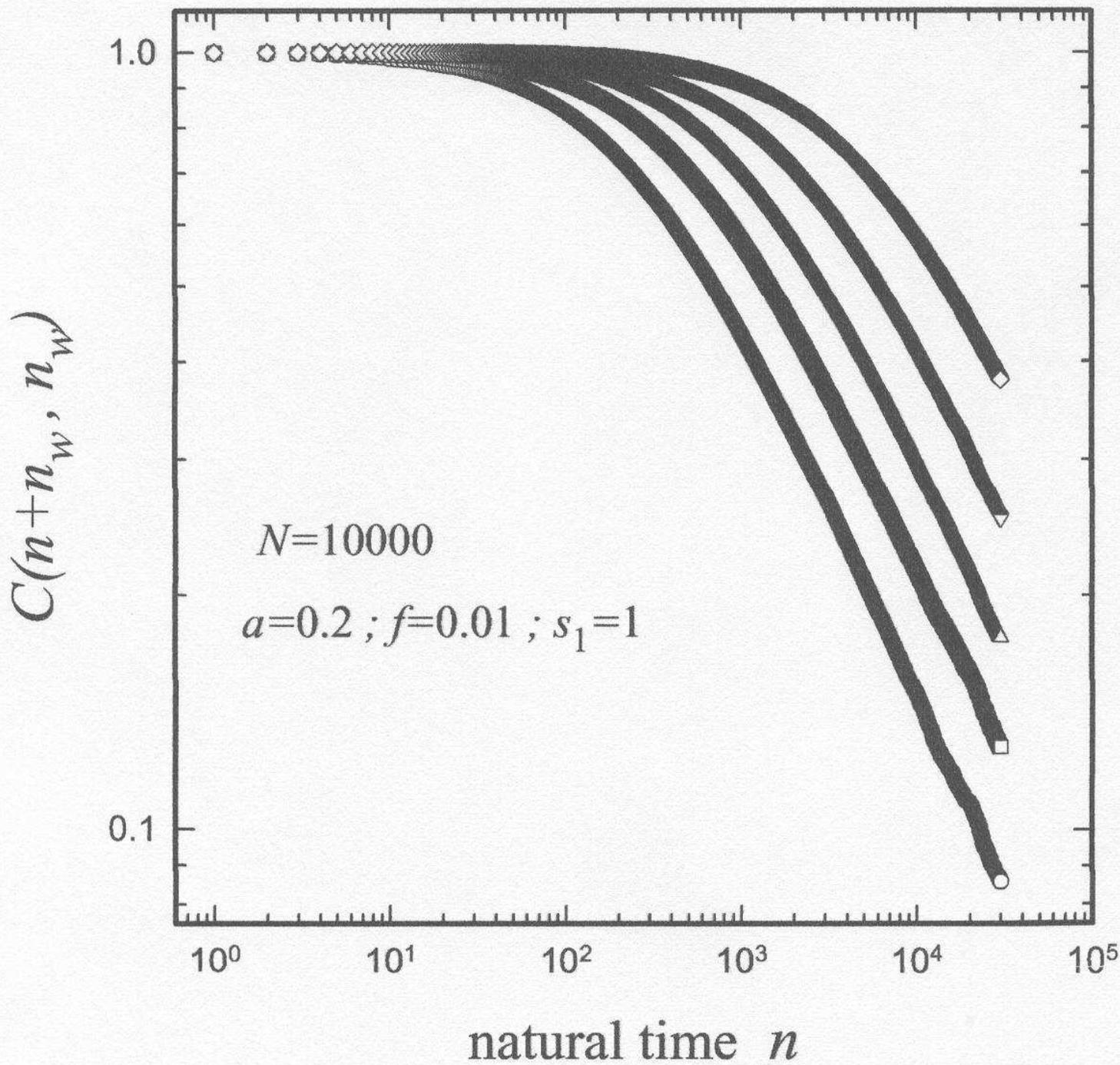
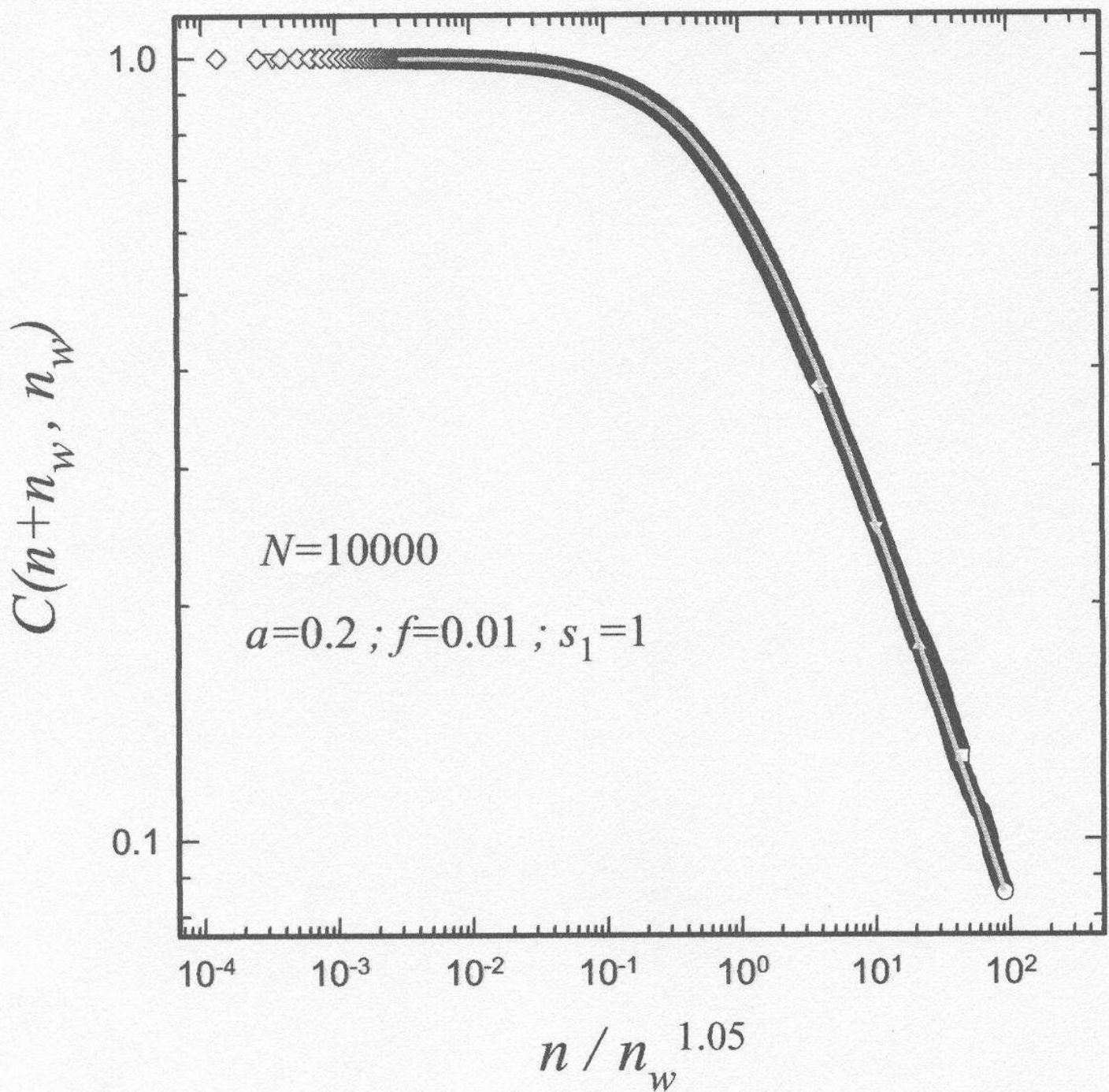
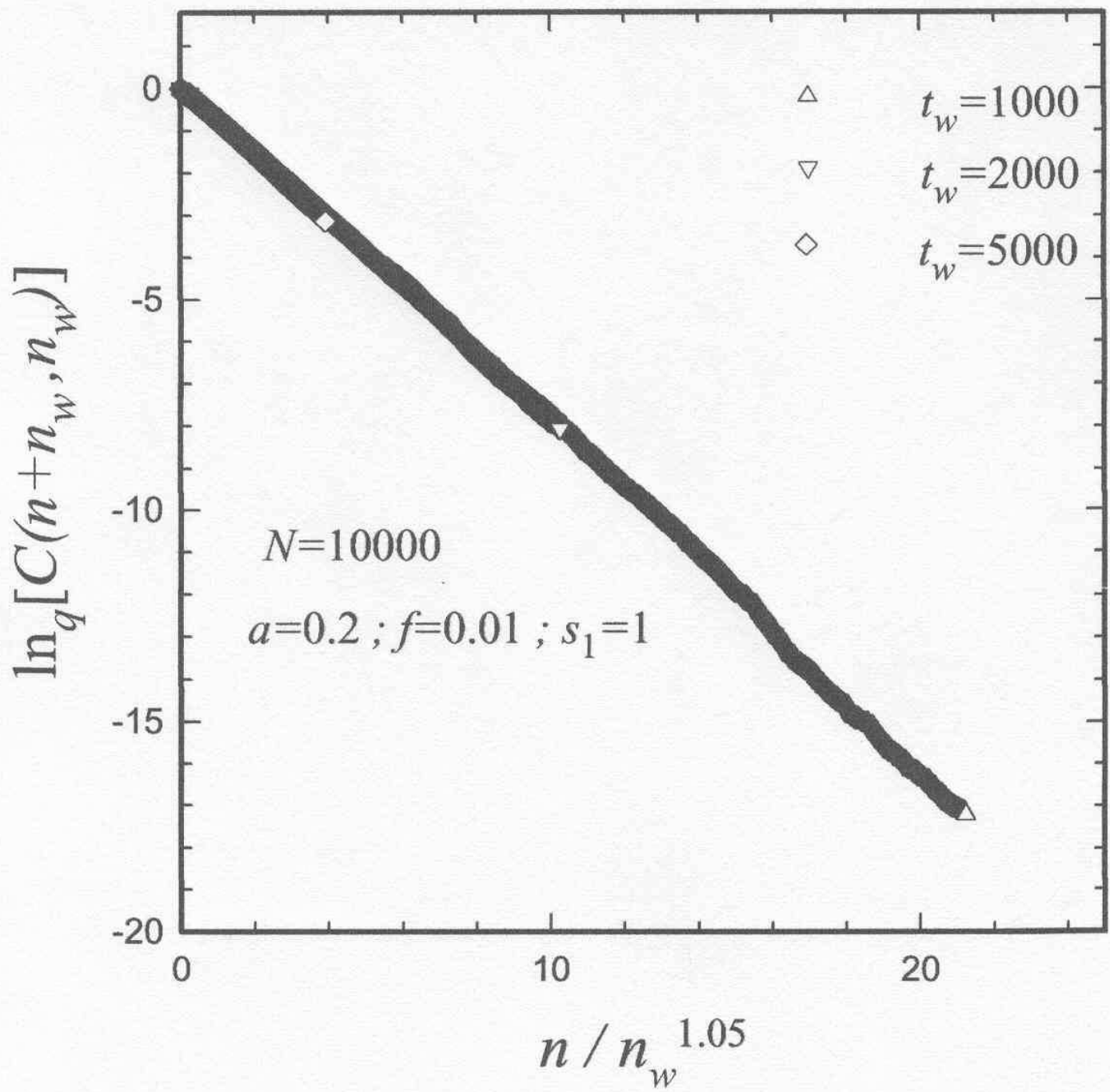
\end{document}